\documentclass[10pt]{article}
\usepackage{graphicx}
\renewcommand{\baselinestretch}{1.4}

\begin{document}
\title{\LARGE \bf \boldmath Search for direct production
of $a_2$(1320) \\ and $f_2$(1270) mesons in $e^+e^-$ annihilation}
\author{ 
\Large M.N. Achasov, S.E. Baru, K.I. Beloborodov, A.V. Berdyugin,  \\
\Large A.G. Bogdanchikov, A.V. Bozhenok, D.A. Bukin, S.V. Burdin, \\
\Large T.V. Dimova, A.A. Drozdetsky, V.P. Druzhinin, \\
\Large M.S. Dubrovin, I.A. Gaponenko, V.B. Golubev, \\
\Large V.N. Ivanchenko, A.A. Korol, M.S. Korostelev, S.V. Koshuba, \\
\Large G.A. Kukartsev, E.V. Pakhtusova, A.A. Polunin, E.E. Pyata, \\
\Large A.A. Salnikov, S.I. Serednyakov, V.A. Sidorov, Z.K. Silagadze, \\
\Large A.N. Skrinsky, V.V. Shary, Yu.M. Shatunov, 
A.V. Vasiljev\footnote{\large 
Fax: +7(3832)342163; e-mail: vasiljev@inp.nsk.su} \\
~~~ \\
\Large{ Budker Institute of Nuclear Physics,} \\
\Large{ Novosibirsk State University,} \\
\Large{ Lavrentiev Avenue, 11,} \\
\Large{ Novosibirsk, 630090, Russia}}
\date{}
\maketitle
\begin{abstract}
\small
A search for direct production of C-even resonances
$a_2$(1320) and $f_2$(1270) in $e^+e^-$ annihilation 
was performed with SND detector at VEPP-2M $e^+e^-$ 
collider. The upper limits of electronic widths of these 
mesons were obtained at 90\,\% confidence level:
\begin{center}                                   
      $\Gamma(a_2(1320) \to e^+e^-) < $0.56\,eV, \\
      $\Gamma(f_2(1270) \to e^+e^-) < $0.11\,eV.
\end{center}
\end{abstract}

\vspace*{5mm}



\renewcommand{\baselinestretch}{1.04}\normalsize
\twocolumn
{\bf 1. Introduction}
\vspace*{3mm}

\begin{figure}[htbp]
\begin{minipage}{0.47\textwidth}
\includegraphics{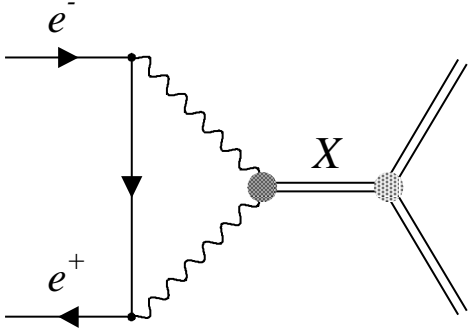}
\caption{ \label{sndlab1}
The diagram of direct production of C-even resonance in
$e^+e^-$ collision.}
\end{minipage}
\end{figure}

Traditional subject of study in $e^+e^-$ collisions are vector 
states with J$^{PC}$=1$^{--}$. Direct production of $C$-even 
mesons (J$^{PC}$=0$^{-+}$, 0$^{++}$, 2$^{++}$, {\ldots} ) is 
also possible via two-photon annihilation (fig.\ref{sndlab1}) 
although it is suppressed by a factor of $\sim \alpha^2$ for 
tensor mesons. Production of  scalar and pseudoscalar states 
is further suppressed by additional 'chirality' factor $m_e^2/s$.
Nevertheless, $e^+e^-$ colliding beam technique remains one of
the most sensitive methods of measurement of electronic widths
of $C$-even resonances ($X$) with masses around 1\,GeV
\cite{vslyaf48,PR91}. 

In the unitarity limit \cite{vslyaf13} when both virtual photons 
(fig.\ref{sndlab1}) are on the mass shell the leptonic width 
is completely determined by imaginary part of the $X \to e^+e^-$ 
transition amplitude which is related to the $X$ two-photon width 
\cite{vslyaf21}. Taking into account both real and imaginary  
parts of the $X \to e^+e^-$ transition amplitude $Z$, the branching 
ratio of $X \to e^+e^-$ decay can be written as follows:

\vspace*{-9mm}
\begin{center}
\begin{eqnarray}
Br(X \to e^+e^-) = \frac{2 \alpha^2}{9} \cdot 
Br(X \to \gamma \gamma) \cdot \nonumber \\ \biggl[1 + 
(\mbox{Re}Z)^2 / (\mbox{Im}Z)^2 \biggr] 
\label{branch}
\end{eqnarray}
\end{center}
At present only branching ratios of pseudoscalar
mesons $\eta \to \mu^+\mu^-$ \cite{PDG} and
$\pi^0 \to e^+e^-$ \cite{pi0} are measured with accuracies 
of 15\,\% and 8\,\% respectively.

The energy range of the electron-positron collider VEPP-2M 
\cite{VEPP} allows to perform a search for production 
of the lightest tensor mesons $f_2(1270)$ and $a_2(1320)$ 
in $e^+e^-$ annihilation. Using the experimental values 
\cite{PDG} of the two-photon widths of these mesons, one can 
estimate their electronic widths in the unitarity limit:

\vspace*{-7mm}
\begin{center}
\begin{equation}
\Gamma(a_2(1320) \to e^+e^-)_{ul} \sim 1 \cdot 10^{-2}\,eV ,
\label{vslequ01}
\end{equation}
\begin{equation}
\Gamma(f_2(1270) \to e^+e^-)_{ul} \sim 3 \cdot 10^{-2}\,eV
\end{equation}
\end{center}

The only experimental attempt to measure these widths was 
taken in the ND experiment at VEPP-2M collider
\cite{ND} in the search for the reactions:

\vspace*{-8mm}
\begin{center}
\begin{equation}
e^+e^- \to a_2(1320) \to \eta \pi^0 ,
\label{vslequ22}
\end{equation}
\begin{equation}
e^+e^- \to f_2(1270) \to \pi^0 \pi^0
\label{vslequ02}
\end{equation}
\end{center}
As a result the following upper limits at 90\,\%
confidence level were obtained:
$\Gamma (a_2(1320) \to e^+e^-) <$ 25\,eV,
$\Gamma (f_2(1270) \to e^+e^-) <$ 1.7\,eV \cite{vslyaf48,PR91}.

\vspace*{7mm}
{\bf 2. Detector and experiment}
\vspace*{3mm}

In the present work the search for the reactions (\ref{vslequ22},
\ref{vslequ02}) was continued.
The experiments \cite{PrBINP9865} were carried out in 1997 and 1999
at VEPP-2M $e^+e^-$ collider with the SND detector~\cite{SND,SND1}.
Four successive scans of the energy range $2E_0$=1.04--1.38\,GeV
with the step $\Delta(2E_0$)=0.01\,GeV were performed.
The total integrated luminosity of 9\,pb$^{-1}$ was
uniformly distributed over this energy range. For present analysis 
only the data with $2E_0$ above 1.15\,GeV 
with an integrated luminosity of 6.5\,pb$^{-1}$ was used.

The SND detector is a universal nonmagnetic detector. Its main
part is a three-layer electromagnetic calorimeter consisted
of 1630 NaI(Tl) crystals covering 90\,\% of 4$\pi$ solid angle.
The energy resolution of the calorimeter for photons with energy
$E$ can be described by the function 
$\sigma_E/E = 4.2\%/\sqrt[4]{E, GeV}$,
the angular resolution is close to 1.5\,$^{\circ}$, 
the resolution over $\pi^0$ invariant mass is approximately 
8\,\%. For measurement of charged particles directions the 
system of two central cylinrical drift chambers is used.
More detailed description of the SND detector can be found
in~\cite{SND1}.

\begin{figure}[htbp]
\begin{minipage}{0.47\textwidth}
\includegraphics{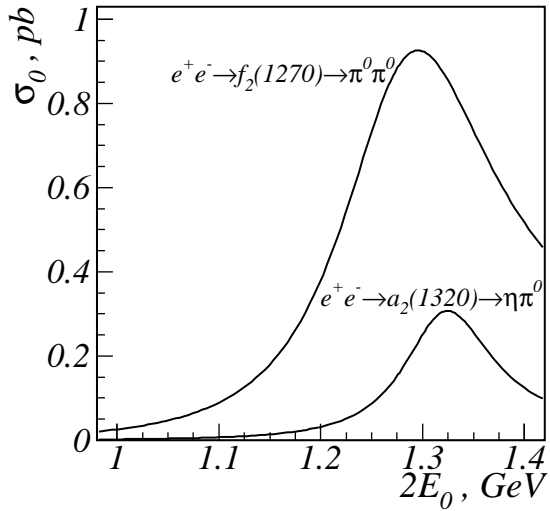}
\caption{ \label{sndlab2}
Energy dependences of the cross sections of the reactions 
$e^+e^- \to f_2(1270) \to \pi^0\pi^0$,
$e^+e^- \to a_2(1320) \to \eta\pi^0$ 
calculated in the unitarity limit.}
\end{minipage}
\end{figure}

A search for the reaction (\ref{vslequ02}) was carried out 
taking into account the differential cross section 
calculated in~\cite{vslyaf40}:

\vspace*{-6mm}
\begin{center}
\begin{eqnarray}
\frac{d\sigma}{d\Omega}=
12.5\cdot\Biggl(\frac{\sqrt{s}}{m}\Biggr)^6\cdot 
\nonumber \\ \frac{\Gamma^2
\cdot B_{ee}\cdot B_{\pi^0\pi^0}}{(m^2-s)^2 + m^2 \Gamma^2}
\cdot sin^2(2\theta) ,
\label{vslequ03}
\end{eqnarray}
\end{center}
where $s= 4E_0^2$; m, $\Gamma$,
B$_{ee}$, and B$_{\pi^0 \pi^0}$ are the $f_2$-resonance mass, 
width, and branching ratios of its decays into $e^+e^-$ and
$\pi^0\pi^0$. The cross sections of the 
reactions (\ref{vslequ22}) and (\ref{vslequ02}), calculated
in the unitarity limit, are shown in the fig.\ref{sndlab2}.
Expected numbers of events, corresponding the collected 
luminosity distribution, are 1 and 4 for the reactions 
(\ref{vslequ22}) and (\ref{vslequ02}) respectively.

\vspace*{7mm}
{\bf 3. Events selection}
\vspace*{3mm}

For the primary selection of events the following cuts 
were applied:

{\it 
\begin{itemize}
\item[ ] four photons and no charged particles are found in an event;
\item[ ] energy deposition in the calorimeter
$E_{tot} > 0.7\cdot(2\cdot E_{beam})$;
\item[ ] total momentum of an event measured by the calorimeter
$P_{tot} < 0.3\cdot(2\cdot E_{beam}/c)$.
\end{itemize}
}
A total of 12.6 thousand events satisfying the above criteria were found.

Main background for the processes (\ref{vslequ22}, \ref{vslequ02}) comes 
from the following reactions with a 3 order of magnitude larger cross 
sections:

\vspace*{-8mm}
\begin{center}
\begin{equation}
e^+e^- \to 4 \gamma~(QED),
\label{vslequ04}
\end{equation}
\end{center}

\vspace*{-10mm}
\begin{center}
\begin{equation}
e^+e^- \to \omega \pi^0 \to \pi^0 \pi^0 \gamma,
\label{omp0n}
\end{equation}
\end{center}
where the reaction (\ref{omp0n}) produces events satisfying  $4\gamma$
selection cuts
due to merging of close photons or 
loss of soft photons through openings in the calorimeter.

Other background processes are the reactions with emission of
hard photon at large angle by one of the initial particles and 
subsequent production of $\rho$, $\omega$, or $\phi$ meson:

\vspace*{-3mm}
\begin{equation}
e^+e^- \to V \gamma \to \pi^0 \gamma \gamma,
~\eta \gamma \gamma,~~V = \rho, \omega, \phi
\label{gammaVeq}
\end{equation}
Their cross sections are 1--2 orders of magnitude 
larger than these of the processes under study
(\ref{vslequ22}, \ref{vslequ02}) \cite{gammaV}.

Additional background comes from the QED processes:

\vspace*{-7mm}
\begin{center}
\begin{equation}
e^+e^- \to 2 \gamma,~3 \gamma
\label{QEDbkg}
\end{equation}
\end{center}
with detection of additional stray photons of beam background. 
Energy spectrum and angular distribution of such photons were
studied on special class of events with trigger from external 
generator. Analysis of these events shows that stray photons 
mainly concentrate at small angles with respect to the beam 
axis and their spectrum decreases sharply with increase of 
energy. To suppress a contribution from the processes 
(\ref{QEDbkg}) with extra photons, the following restrictions 
on angle $\theta_{ \gamma}$ and energy {\it E}$_{ \gamma}$ of 
each photon in the event were applied:
{\it 
\begin{itemize}
\item[ ] 27\,$^{\circ} < \theta_{ \gamma} <$ 153\,$^{\circ}$, 
\hspace*{1mm} E$_{ \gamma} >$ 0.1$\cdot$E$_{beam}$.
\end{itemize}
}
Although these cuts reduce efficiency for the processes
under study (\ref{vslequ22}, \ref{vslequ02}) by 30\,\%,
they strongly, by about five times, suppress contribution
of the QED processes (\ref{vslequ04}, \ref{QEDbkg}). After 
all above listed cuts 2036 events were selected, which
correspond to the total detection cross section $\sim$ 0.3\,nb.

To suppress background events with merged photons the special 
parameter $\zeta$~\cite{XINM} was used. This parameter is 
a measure of likelihood of the hypothesis that given transverse
energy deposition profile of a photon  cluster in the calorimeter 
can be attributed to a single photon emitted from the beam 
interaction point. The requirement
{\it 
\begin{itemize}
\item[ ] $\zeta < $0
\end{itemize}}
for all photons in an event allows to suppress significantly 
the contribution of events with merged photons and events 
of the process
\vspace*{-7mm}
\begin{center}
\begin{equation}
e^+e^- \to K_S K_L \to \pi^0 \pi^0 K_L 
\label{KSKL}
\end{equation}
\end{center}
with nuclear interaction of K$_L$. This cut reduces the number 
of experimental events by 40\,\% while the detection efficiencies 
for the processes (\ref{vslequ22}) and (\ref{vslequ02}) decrease 
by only 6\,\% and 4\,\% respectively.

For the events left, kinematic fit with requirement of
energy-momentum conservation was performed and corresponding value 
of $\chi^2$ was calculated. For further analysis 842 events with
{\it 
\begin{itemize}
\item[ ] $\chi^2 < $20
\end{itemize}}
were selected. This number is in a good agreement with expected 
contribution of the background processes (\ref{vslequ04} - 
\ref{gammaVeq}) obtained by simulation. About 65\,\% of it comes 
from the process (\ref{omp0n}).

\begin{figure}[htbp]
\begin{minipage}{0.47\textwidth}
\includegraphics{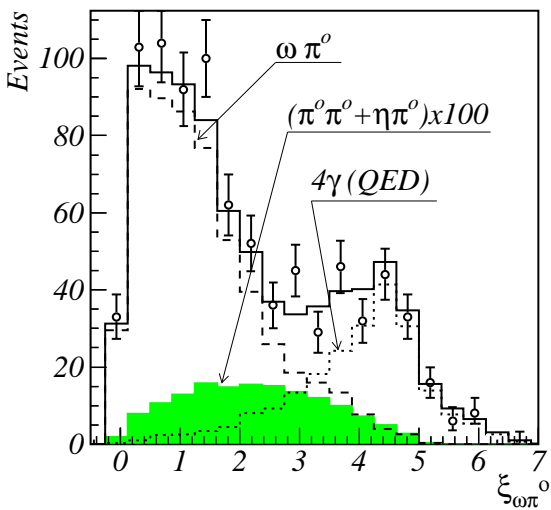}
\caption{ \label{sndlab3}
Distribution of 842 experimental events over the parameter
$\xi_{\omega\pi^0}$ (circles with error bars). 
The clear histogram is the sum of expected contributions
of the main background processes (\ref{vslequ04} -- \ref{gammaVeq}),
shaded histogram is an expected signal of the processes
under study (\ref{vslequ22}, \ref{vslequ02}) multiplied by
100. Dashed and dotted lines show individual contributions of
the main background processes
(\ref{omp0n}) and (\ref{vslequ04}) respectively.}
\end{minipage}
\end{figure}

To suppress contribution of the process (\ref{omp0n}) the special 
parameter $\xi_{\omega\pi^0}$ was constructed taking into account 
the topology of the process (\ref{omp0n}) with only four photons 
detected. Three hypotheses were considered:\\
1) undetected photon is from recoil $\pi^0$ (12 possible 
combinations);\\
2) undetected photon is the recoil photon from $\omega$ decay
(6 possible combinations);\\
2) undetected photon comes from $\pi^0$ in $\omega$ decay 
(6 possible combinations).\\
For all 24 possible combinations of photons the values of

\vspace*{-7mm}
\begin{center}
\begin{equation}
\chi^2_i = \frac{(m_1-m_{\omega})^2}{\sigma_{\omega}^2}+
\frac{(m_2-m_{\pi^0})^2}{\sigma_{\pi^0}^2}
\label{xop1a}
\end{equation}
\end{center}
were calculated. Here $m_1$, depending on the combination being 
considered, is either invariant mass of three photons or recoil 
mass of two photons, $m_2$ is an invariant mass of two photons 
for a given combination; $m_{\omega}$ and $m_{\pi^0}$ --- 
the masses of $\omega$ and $\pi^0$ mesons respectively. 
The parameter $\xi_{\omega\pi^0}$ was defined as 

\vspace*{-7mm}
\begin{center}
\begin{equation}
\xi_{\omega\pi^0} = ln ( 1 + {min}_i (\chi^2_i)),~~i = 1, \ldots, 24.
\label{xop1b}
\end{equation}
\end{center}
It is seen from $\xi_{\omega\pi^0}$ distributions in the 
fig.\ref{sndlab3} that events of the process (\ref{omp0n}) 
concentrate in the left side of the plot while processes under
study (\ref{vslequ22}, \ref{vslequ02}) have flatter and wider 
spectrum allowing the use of this parameter for further cuts.

\begin{figure}[htbp]
\begin{minipage}{0.47\textwidth}
\includegraphics{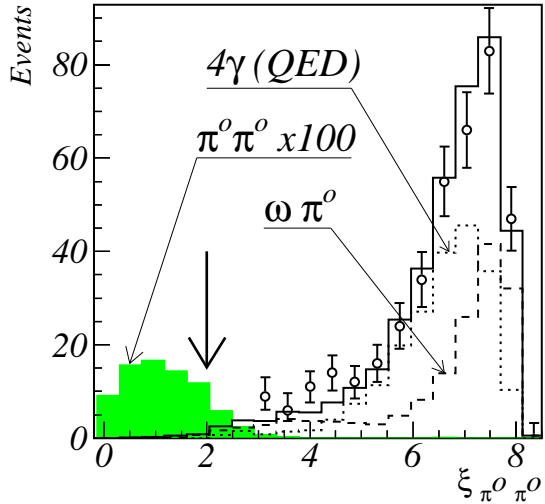}
\caption{ \label{sndlab4}
The $\xi_{\pi^0\pi^0}$ distributions:
circles with error bars --- experimental data; clear
histogram --- expected contribution of the background
processes (\ref{vslequ04} - \ref{gammaVeq}), shaded
histogram --- expected signal of the process (\ref{vslequ02})
multiplied by 100. In addition the individual contributions of
main backgrounds, (\ref{omp0n}) --- dashed line and
(\ref{vslequ04}) --- dotted line, are shown separately.
Arrow shows the cut $\xi_{\pi^0\pi^0}<$2 used for final selection.}
\end{minipage}
\end{figure}

\begin{figure}[htbp]
\begin{minipage}{0.47\textwidth}
\includegraphics{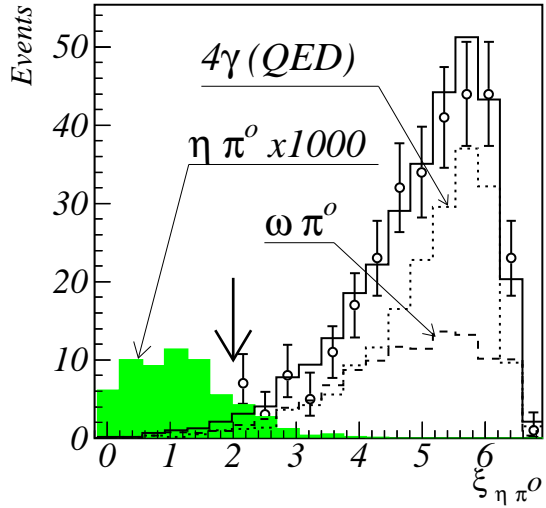}
\caption{ \label{sndlab5}
The $\xi_{\eta\pi^0}$ distributions:
circles with error bars --- experimental data; clear
histogram --- expected contribution of the background
processes (\ref{vslequ04} - \ref{gammaVeq}), shaded
histogram --- expected signal of the process (\ref{vslequ02})
multiplied by a factor 1000. In addition, individual contributions 
of main backgrounds, (\ref{omp0n}) --- dashed line --- and
(\ref{vslequ04}) --- dotted line, are shown separately.
Arrow shows the cut $\xi_{\eta\pi^0}<$2 used for final selection.}
\end{minipage}
\end{figure}

Similarly to eq.~\ref{xop1a},\ref{xop1b} the parameters 
$\xi_{\eta\pi^0}$, $\xi_{\pi^0\pi^0}$ and $\xi_{\omega\pi^0\gamma}$, 
$\xi_{\omega\eta\gamma}$, $\xi_{\phi\eta\gamma}$, 
$\xi_{\phi\pi^0\gamma}$ for selection of the processes 
(\ref{vslequ22}), (\ref{vslequ02}) and (\ref{gammaVeq}) respectively 
were constructed.

\vspace*{7mm}
{\bf 4. Final events selection and results}
\vspace*{3mm}

To select candidate events of the reaction (\ref{vslequ02}) 
the following cuts were imposed:

\vspace*{-7mm}
\begin{equation}
\hspace*{2mm} \xi_{\omega\pi^0} > 1.5, \hspace*{1mm}
\hspace*{2mm} \xi_{\omega\pi^0\gamma} > 2, \hspace*{1mm}
\hspace*{2mm} \xi_{\pi^0\pi^0} < 2.
\label{selxpp}
\end{equation}
The $\xi_{\pi^0\pi^0}$ distribution before the last cut from 
(\ref{selxpp}) is shown in fig.~\ref{sndlab4}. No experimental events 
passed selection cuts, while 0.7 events of the process under study 
(\ref{vslequ02}) (in the unitarity limit) and 1.4 events of the 
background processes are expected from simulation. The selection 
efficiency for the process (\ref{vslequ02}) is close to 17\,\%.

To search for the reaction (\ref{vslequ22}) the following cuts 
were applied to 842 experimental events selected above:

\vspace*{-7mm}
\begin{equation}
\hspace*{2mm} \xi_{\omega\pi^0} > 2, \hspace*{1mm}
\hspace*{2mm} \xi_{\omega\eta\gamma} > 2, \hspace*{1mm}
\hspace*{2mm} \xi_{\phi\eta\gamma} > 1, \hspace*{1mm}
\hspace*{2mm} \xi_{\eta\pi^0} < 2.
\label{selxpe}
\end{equation}
The $\xi_{\eta\pi^0}$ distribution before the last cut from 
(\ref{selxpe}) is shown in fig.~\ref{sndlab5}. Here again no 
experimental events were found. Expected number of events of 
the process under study (\ref{vslequ22}) is 0.05, the selection 
efficiency is about 5\,\% taking into account all decay modes 
of $\eta$ meson. Calculated contribution of the background 
processes (\ref{vslequ04} -- \ref{gammaVeq}) is 5.5 events.

\begin{table*}
\begin{center}
\begin{minipage}{0.76\textwidth}
\caption{\label{TablSumm}
The upper limits of branching ratios and electronic widths
of tensor mesons $a_2$(1320) and $f_2$(1270) obtained in 
this work as compared with current experimental data and
theoretical calculations in the unitarity limit ~\cite{vslyaf13}.}

\vspace*{3mm}
{\renewcommand\baselinestretch{1.2}
\begin{tabular}{|{l}|{c}|{c}|{c}|}
\hline
& this work
& ND'91 \cite{PR91}
& calculation in the \\
& (SND'2000)
& (PDG'98 \cite{PDG})
& unitarity limit ~\cite{vslyaf13} \\
\hline
$ Br(a_2 \to e^+e^-)$        & $<$ 6 $\cdot$ 10$^{-9} $ 
& $<$ 2.3 $\cdot$ 10$^{-7} $ & 1.1 $\cdot$ 10$^{-10} $ \\
\hline
$ Br(f_2 \to e^+e^-)$        & $<$ 6 $\cdot$ 10$^{-10} $     
& $<$ 9 $\cdot$ 10$^{-9} $   & 1.6 $\cdot$ 10$^{-10} $ \\
\hline
$\Gamma(a_2 \to e^+e^-)$, eV & $<$ 0.56  & $<$ 25  & 0.012 \\
\hline
$\Gamma(f_2 \to e^+e^-)$, eV & $<$ 0.11  & $<$ 1.7 & 0.029 \\
\hline
\end{tabular}
}
\end{minipage}
\end{center}
\end{table*}

We used the following formulae to obtain the upper limits
of branching ratios of electronic decays of tensor mesons
$T$ = $a_2$(1320), $f_2$(1270):

\vspace*{-3mm}
\begin{equation}
Br ( T \to e^+e^-) < \frac{k_0}{N_{expt}}~.
\label{UppLim}
\end{equation}
Here $N_{expt} = \sum \Delta L(E_i) \cdot \sigma(E_i) \cdot 
(1+\delta) \cdot \epsilon(E_i)$ is expected number of events 
of the process under study with $Br(T \to e^+e^-)=1$ , $k_0$=2.44 
is a 90\,\% CL Poisson limit in case of none experimental events 
observed (~\cite{PDG}, p.~177), $\Delta L(E_i)$ is an integrated 
luminosity at the energy $E_i$, $\sigma(E_i)$ and $\epsilon(E_i)$ 
are the cross section and the selection efficiency of the process 
under study, calculated by MC simulation, $\delta$ is a radiative 
correction. In the table (\ref{TablSumm}) upper limits at 90\,\% CL 
on branching ratios and electronic widths of $a_2$(1320) and 
$f_2$(1270) calculated according to eq.(\ref{UppLim}) as well as 
theoretical predictions are presented.

\vspace*{7mm}
{\bf 5. Discussion}
\vspace*{3mm}

The upper limits of the $a_2 \to e^+e^-$ and $f_2 \to e^+e^-$
branching ratios obtained in this work are respectively 45 and 15 
times lower than previous experimental values \cite{vslyaf48}.
Our limit for the electronic width of $f_2$(1270) is only four 
times higher than its unitarity limit. It allows for the first 
time to place a meaningful experimental limit for the ratio of the 
real and imaginary parts of $f_2 \to e^+e^-$ transition amplitude:
\vspace*{2mm}
\begin{equation}
(\mbox{Re}Z)^2 / (\mbox{Im}Z)^2 < 2.8 
\label{FFLim}
\end{equation}
at 90\,\% confidence level. Unfortunately, we could not find any 
theoretical estimates for this parameter. To observe the process 
(\ref{vslequ02}) with SND detector it is necessary to increase 
integrated luminosity by 1--2 orders of magnitude.

\vspace*{7mm}
{\bf Acknowledgments}
\vspace*{3mm}

This work is supported in part by Russian Fund for Basic Researches, 
grants No.~00-02-17481 and 00-15-96802, and STP ``Integration'' No.~274.

\end{document}